\documentclass[12pt]{article}
\usepackage{amsmath,bbm,amssymb}
\usepackage{amsthm}
\usepackage{comment}
\usepackage{graphicx, graphics}
\usepackage{textcomp}
\usepackage{amsfonts}

\usepackage{psfrag,pst-node,subfigure,rotating, amsmath, bbm, amsthm, amssymb, amsthm, setspace, picture, epsfig, amsfonts, upgreek}

\usepackage{multirow}

\usepackage{arydshln}
\usepackage{algorithm}
\usepackage{algorithmic}

\usepackage{arydshln}

\newtheorem{theorem}{Theorem}
\newtheorem{lemma}{Lemma}

\def\v{\mathbf}\def\m{\mathbf}\def\vg{\boldsymbol}\def\s{\mathcal}\def\C{{\mathrm{Cov}}}

\begin{document}
\begin{center}
    {\Large\bf Design and analysis of computer experiments with both numeral and distribution inputs}
\\[2mm] Chunya Li$^{1}$, Xiaojun Cui$^{2}$, Shifeng Xiong$^{3}$\footnote{Corresponding author, Email: xiong@amss.ac.cn}
		{\footnotesize\\ 1. School of Mathematics, Physics, and Statistics, Shanghai University of Engineering Science\\ Shanghai 201620, China
\\[1mm] 2. Department of Mathematics, Nanjing University\\ Nanjing 210093, China
\\[1mm] 3. NCMIS, KLSC, Academy of Mathematics and Systems Science, Chinese Academy of Sciences\\ Beijing 100190, China}	
\end{center}

\vspace{1cm} \noindent{\bf Abstract}\quad Nowadays stochastic computer simulations with both numeral and distribution inputs are widely used to mimic complex systems which contain a great deal of uncertainty. This paper studies the design and analysis issues of such computer experiments. First, we provide preliminary results concerning the Wasserstein distance in probability measure spaces. To handle the product space of the Euclidean space and the probability measure space, we prove that, through the mapping from a point in the Euclidean space to the mass probability measure at this point, the Euclidean space can be isomorphic to the subset of the probability measure space, which consists of all the mass measures, with respect to the Wasserstein distance. Therefore, the product space can be viewed as a product probability measure space. We derive formulas of the Wasserstein distance between two components of this product probability measure space. Second, we use the above results to construct Wasserstein distance-based space-filling criteria in the product space of the Euclidean space and the probability measure space. A class of optimal Latin hypercube-type designs in this product space are proposed. Third, we present a Wasserstein distance-based Gaussian process model to analyze data from computer experiments with both numeral and distribution inputs. Numerical examples and real applications to a metro simulation are presented to show the effectiveness of our methods.


\vspace{1cm} \noindent{{\bf KEY WORDS:} Gaussian process model; metro simulation; mixed inputs; space-filling design; Wasserstein distance.}



\section{Introduction}\label{sec:intro}

Nowadays more and more real systems can be studied virtually by means of computer codes. Since it is often time-consuming to run such codes, elaborate design and modeling for computer experiments are necessary. A great amount of literature has studied statistical issues related to computer experiments, including experimental design, surrogate model, optimization, and many others (Fang, Li, and Sudjianto 2005; Santner, Williams, and Notz 2018). Most existing researches focus on deterministic computer experiments, i.e., the simulator produces the same result if run twice using the same set of inputs. Simulations for physical phenomena through numerically solving mathematical
models (differential equations) are usually deterministic. However, many real complex systems contain a great deal of uncertainty, and stochastic computer simulations can better mimic them.
If run twice using the same set of inputs, a stochastic computer simulation yields different results. In other words, its output is a probability distribution. Typical stochastic computer simulations can be found in urban transportation (Elefteriadou 2014) since the movements of vehicles, pedestrians, and passengers have high degree of randomness. Other examples of stochastic computer simulations appear in physical phenomenon simulations with random inputs (Xiu 2010), reliability (Nanty, Helbert, Marrel et al. 2016), and social science (Squazzoni, Jager, and Edmonds 2014).

The randomness of the output is caused by random numbers generated from some stochastic mechanism involved in the computer codes. The probability distributions behind these random number are actually inputs of the simulation. Therefore, stochastic computer simulations can be viewed as those having distribution inputs. There are a few papers on the design and modeling issues of computer experiments with function inputs (Muehlenstaedt, Fruth, and Roustant 2017; Tan 2019; Betancourt, Bachoc, Kleina et al. 2020; Chen, Mak, Joseph et al. 2021), but very limited on distribution inputs. Bachoc, Gamboa, Loubes et al. (2018) constructed the Wasserstein distance-based Gaussian process model with a one-dimensional distribution input. Bachoc, Suvorikova, Ginsbourger et al. (2020) discussed such a model with multidimensional distribution inputs. Note that actual stochastic simulations often have mixed types of inputs. In this paper we focus on computer experiments with both numeral and distribution inputs. To the best of our knowledge, the design and analysis issues of such computer experiments have not been investigated in the literature.

In a probability measure space, the Wasserstein distance, which is related to the optimal transport problem (Pamaretps and Zemel 2020), possesses a number of good mathematical properties. It thus has been widely applied in machine learning and statistics (Arjovsky, Chintala, and Bottou 2017; Peyr\'{e} and Cuturi 2019), including the construction of surrogate models of computer simulations mentioned above. For the mixed-input problem we face, we find that it is also very suitable. We prove that, through the mapping from a point in the Euclidean space to the mass probability measure at this point, the Euclidean space can be isomorphic to the subset of the probability measure space, which consists of all the mass measures, with respect to the Wasserstein distance. Therefore, the product space of the Euclidean space and the probability measure space can be viewed as a product probability measure space. By this way, the mixed inputs are unified within this product probability measure space. We derive formulas of the Wasserstein distance between two components of the product space, and use it to define Wasserstein distance-based space-filling criteria. A class of discretization-based approximate maximin designs in a one-dimensional probability measure space and a class of Latin hypercube-type designs in the product space are proposed. For the modeling issue, we focus on real-valued responses which are numeral features such as the expectation of the output distribution.
We use the Wasserstein distance to construct a Gaussian process model with mixed inputs, and discuss the corresponding estimation and prediction methods. Numerical experiments with test functions are presented to evaluate our design and prediction methods. Real applications to a metro passenger flow simulation are provided. A metro system contains a great deal of uncertainty caused by passengers' uncertain movements. We apply the proposed methods to build the surrogate model of the simulation, which shows how passengers' travel times depend on their walking time distribution and a boarding probability parameter.

The rest of this paper is organized as following. Section \ref{sec:wd} gives preliminary results concerning the Wasserstein distance. Section \ref{sec:sd} constructs experimental designs in the product space of the Euclidean space and the probability measure space. Section \ref{sec:gp} discusses the Gaussian process model for the mixed inputs. Section \ref{sec:experiment} shows numerical results with test functions, Section \ref{sec:real} provides real applications to the metro simulation. Section \ref{sec:dis} concludes the paper with a discussion. Technical proofs are given in the Appendix.

\section{The Wasserstein distance}\label{sec:wd}

For $p,q \geqslant1$ and positive integer $d$, consider the set $\mathcal{P}_{q,p}(\mathbb{R}^d)$ of probability measures on $\mathbb{R}^d$ with a finite moment (with respect to the $\ell_p$ norm) of order $q$. For $\mu,\nu\in\mathcal{P}_{q,p}(\mathbb{R}^d)$, we denote by $\Pi(\mu,\nu)$ the set of all probability measures $\pi$ over the product set $\mathbb{R}^d\times\mathbb{R}^d$ with first (resp. second) marginal $\mu$ (resp. $\nu$). Any element in  $\Pi(\mu,\nu)$ is called a coupling measure of $\mu$ and $\nu$.  The transportation cost with cost  $\ell^q_p$  between\ $\mu$\ and\ $\nu$ is defined as $$\mathcal{T}_{q,p}(\mu,\nu)=\inf_{\pi\in\Pi(\mu,\nu)}\int\|\v{x}-\v{y}\|_p^q d\pi(\v{x},\v{y}),$$ where $\|\cdot\|_p$ denotes the $\ell_p$ norm.
Since  $\mathbb{R}^d$, equipped with $\ell_p$ norm, is a Polish space, it is well known that the above infimum can be reached. The Wasserstein distance (also called Monge-Kantorovich distance) with ground metric $\ell_p$ between $\mu$ and $\nu$ is defined as$$W_{q,p}(\mu,\nu)=\mathcal{T}_{q,p}(\mu,\nu)^{1/q}.$$ For $d=1$ and $p=q$, the Wasserstein distance can be computed by (Villani 2009)
\begin{equation}\label{wdc}W_{p,p}(\mu,\nu)=\left\{\int_0^1\left|F_\mu^{-1}(t)-F_\nu^{-1}(t)\right|^pdt\right\}^{1/p},\end{equation}where $F_\mu$ and $F_\nu$ are the cumulative distribution functions of $\mu$\ and\ $\nu$, respectively, and $F_\mu^{-1}(t)=\inf\{u:\ F_\mu(u)\geqslant t\}$, $F_\nu^{-1}(t)=\inf\{u:\ F_\nu(u)\geqslant t\}$.

Let $\delta_{\v{x}}$ denote the mass probability measure at $\v{x}\in\mathbb{R}^d$, and $\Delta^d=\{\delta_{\v{x}}:\ \v{x}\in\mathbb{R}^d\}$. Consider the product space $\Delta^d\times\mathcal{P}_{q,p}(\mathbb{R})$, which is a subset of $\mathcal{P}_{q,p}(\mathbb{R}^{d+1})$.

\begin{lemma}\label{lemma:ei}For $\v{x},\v{y}\in\mathbb{R}^d$, we have\begin{equation*}W_{q,p}(\delta_{\v{x}},\delta_{\v{y}})
=\|\v{x}-\v{y}\|_p.\end{equation*}\end{lemma}

This lemma indicates that, with the mapping $\v{x}\mapsto\delta_{\v{x}}$, $\mathbb{R}^d$ (equipped with the $\ell_p$ distance) is isomorphic to $\Delta^d$ with respect to the Wasserstein distance. Therefore, the product space $\mathbb{R}^d\times\mathcal{P}_{q,p}(\mathbb{R})$ is isomorphic to $\Delta^d\times\mathcal{P}_{q,p}(\mathbb{R})$, which is a subspace of $\mathcal{P}_{q,p}(\mathbb{R}^{d+1})$. We view any combination of numeral and distribution inputs as a component of this probability measure space. This provides a way to unify such mixed inputs in $\mathbb{R}^d\times\mathcal{P}_{q,p}(\mathbb{R})$.

For the special product probability measure space $\Delta^d\times\mathcal{P}_{q,p}(\mathbb{R})$, the Wasserstein distance between its two components has the following expression.
\begin{theorem}\label{th:wf}For $\v{x},\v{y}\in\mathbb{R}^d$, $\mu,\nu\in\mathcal{P}_{q,p}(\mathbb{R})$, we have
\begin{equation*}W_{q,p}(\delta_{\v{x}}\times\mu,\delta_{\v{y}}\times\nu)=\left\{\int_0^1  c(F_\mu^{-1}(t)-F_\nu^{-1}(t)) dt\right\}^{1/q},\end{equation*}
where $c(z) = (\|\v{x}-\v{y}\|_p^p+|z|^p)^{\frac{q}{p}}$.
\end{theorem}

This theorem presents formula for calculating the Wasserstein distance in the space we focus on. In particular, we provide its two important special cases, \begin{eqnarray}&&W_{1,1}(\delta_{\v{x}}\times\mu,\delta_{\v{y}}\times\nu)=\|\v{x}-\v{y}\|_1+W_{1,1}(\mu,\nu),\label{w11}
\\&&W_{2,2}(\delta_{\v{x}}\times\mu,\delta_{\v{y}}\times\nu)^2=\|\v{x}-\v{y}\|_2^2+W_{2,2}(\mu,\nu)^2.\label{w22}\end{eqnarray}

\section{Experimental design}\label{sec:sd}

From the results in the previous section, the Wasserstein distance can be viewed as an extension of the $\ell_p$ distance in the Euclidean space to the probability measure space. Similar to the Euclidean space, we can define the Wasserstein distance-based space-filling designs in the probability measure space. In this section we first present a method to construct space-filling designs in the space of probability measures in one dimension. We then introduce Latin Hypercube (LH)-type space-filling designs for the product space of the Euclidean space and the probability measure space.

\subsection{Space-filling designs in $\mathcal{P}_{q,p}([0,1],\tau)$}\label{subsec:sdw}

Note that distribution inputs of stochastic computer simulations usually have compact supports and certain degrees of smoothness. We consider the space $\mathcal{P}_{q,p}([0,1],\tau)=\{\mu\in\mathcal{P}_{q,p}(\mathbb{R}):\ \text{support of}\ \mu \subset[0,1],\ \sup_{x,y\in[0,1],\ x\neq y}|F_\mu(x)-F_\mu(y)|/|x-y|\leqslant\tau\}$ for some constant $\tau>0$. For a design of $n$ runs, $\mathcal{D}=\{\mu_1,\ldots,\mu_n\}\subset\mathcal{P}_{q,p}([0,1],\tau)$, the minimum Wasserstein distance criterion is
\begin{equation}\label{mc} \mathrm{mdc}(\mathcal{D})=\min_{1\leqslant i<j\leqslant n}W_{q,p}(\mu_i,\mu_j).\end{equation}Note that the parameter $q$ in the Wasserstein distance can be arbitrary due to finite support. We can compute $W_{q,p}(\mu_i,\mu_j)$ by \eqref{wdc}. The design that maximizes the criterion \eqref{mc} is defined as the maximin Wasserstein distance design. Similarly we can define minimax Wasserstein distance design and minimum $\phi_p$ design, which are analogues of those in Euclidean space (Johnson, Moore, and Ylvisaker 1990; Morris and Mitchell 1995).

\renewcommand{\algorithmicrequire}{\textbf{Input}:}
\renewcommand{\algorithmicensure}{\textbf{Steps}:}
\floatname{algorithm}{Algorithm}
\begin{algorithm}[thb]
\caption{\label{ag:cd}\quad The block coordinate descent algorithm}
\begin{algorithmic}[1]
\REQUIRE ~~\\ $n,\ m,\ \tau,\ \varepsilon$. \ENSURE ~~\\\STATE \textbf{Initialization:} Select $\v{t}_1^{(0)},\ldots,\v{t}_n^{(0)}\in E_{m-1}(\tau)$. \STATE \textbf{Iteration:} For each $k=0,1,\ldots$,
\\ for $i=1,\ldots,n$, solve\begin{equation*}\begin{split}&\v{t}^{(k+1)}_i=\arg\max_{\v{t}_i} \xi\left(\v{t}_1^{(k+1)},\ldots,\v{t}_{i-1}^{(k+1)},\v{t}_i,\v{t}_{i+1}^{(k)},\ldots,\v{t}_n^{(k)}\right),
\\&\text{subject to}\quad \v{t}_i\in E_{m-1}(\tau),\end{split}\end{equation*}
If $\xi\left(\v{t}_1^{(k+1)},\ldots,\v{t}_n^{(k+1)}\right)-\xi\left(\v{t}_1^{(k)},\ldots,\v{t}_n^{(k)}\right)<\varepsilon$, then stop the iterations, and output $\mathcal{D}=\{\mu_1,\ldots,\mu_n\}$ corresponding to $\v{t}_1^{(k+1)},\ldots,\v{t}_n^{(k+1)}$. \\ Otherwise, $k\leftarrow k+1$.
\end{algorithmic}
\end{algorithm}

Since $\mathcal{P}_{q,p}([0,1],\tau)$ is an infinity-dimensional space, the problem in optimizing \eqref{mc} is difficult. We use a discretization method to approximate it.
For $\mu\in\mathcal{P}_{q,p}([0,1],\tau)$, $F_\mu$ can be approximated by the piecewise  linear function \begin{eqnarray}&&\tilde{F}_\mu(x)\nonumber\\&=&\sum_{i=1}^{m-1}\left\{F_\mu\left(\frac{i-1}{m-1}\right)+(m-1)\left(x-\frac{i-1}{m-1}\right)
\left[F_\mu\left(\frac{i}{m-1}\right)-F_\mu\left(\frac{i-1}{m-1}\right)\right]\right\}\nonumber\\&&\quad\quad \cdot I\left(x\in\left[\frac{i-1}{m-1},\frac{i}{m-1}\right]\right),\label{tf}\end{eqnarray} where $I$ is the indicator function. Therefore, $\mathcal{P}_{q,p}([0,1],\tau)$ can be approximated by the finite dimensional space \begin{eqnarray}&&\widetilde{\mathcal{P}}_{q,p,m-1}([0,1],\tau)\nonumber\\&=&\Big\{\mu:\ F_\mu(x)=\sum_{i=1}^{m-1}\left[s_i+(m-1)\left(x-\frac{i-1}{m-1}\right)
\left(s_{i+1}-s_i\right)\right]I\left(x\in\left[\frac{i-1}{m-1},\frac{i}{m-1}\right]\right),\nonumber\\\nonumber&&\ \ \text{for all}\ (s_1,\ldots,s_m)'\ \text{with}\ 0=s_1\leqslant s_1\leqslant\cdots\leqslant s_m=1,\ s_{i+1}-s_i\leqslant \tau/(m-1),\nonumber\\&&\quad i=1,\ldots,m-1\Big\}.\label{ap}\end{eqnarray} Each $\mu\in\widetilde{\mathcal{P}}_{q,p,m-1}([0,1],\tau)$ corresponds to the vector $(s_1,\ldots,s_m)'$ of $m-2$ degrees of freedom. Let $t_i=s_{i+1}-s_{i},\ i=1,\ldots,m-1$. Then $\mu$ is specified by the vector $(t_1,\ldots,t_{m-1})'\in E_{m-1}(\tau)=\{(x_1,\ldots,x_{m-1})':\ 0\leqslant x_i\leqslant\tau/(m-1),\ i=1,\ldots,m-1,\ \sum_{i=1}^{m-1}x_i=1\}$. The maximin Wasserstein distance design problem reduces to \begin{eqnarray}&&\max\ \ \xi(\v{t}_1,\ldots,\v{t}_{n})=\min_{1\leqslant i<j\leqslant n}W_{q,p}(F_i,F_j)\label{pro}
\\&&\ \ \text{subject to}\quad \v{t}_1,\ldots,\v{t}_{n}\in E_{m-1}(\tau).\nonumber\end{eqnarray}Let $\{\v{t}_1^*,\ldots,\v{t}_n^*\}$ denote the solution to \eqref{pro}. Then the corresponding $\{\mu_1^*,\ldots,\mu_n^*\}$ can be viewed as an (approximate) maximin Wasserstein distance design in $\mathcal{P}_{q,p}([0,1],\tau)$.

\begin{figure}[t]
	\begin{center}
		\scalebox{0.7}[0.7]{\includegraphics{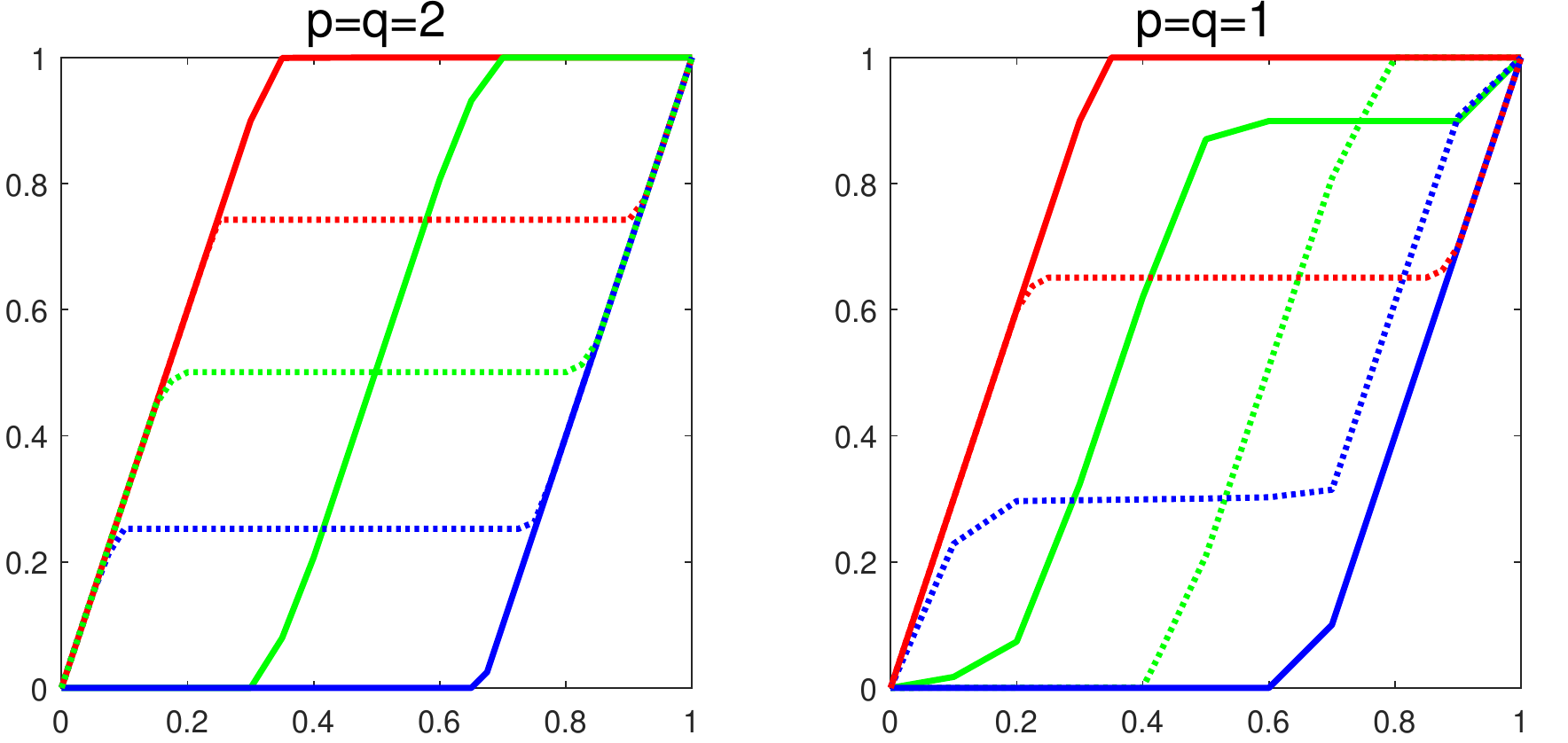}}
	\end{center}
	\caption{maximin $W_{q,p}$ distance designs in $\mathcal{P}_{q,p}([0,1],3)$ for $n=6$.}\label{fig:n6}
\end{figure}

There are $n\times(m-1)$ variables we need to optimize in the approximate problem \eqref{pro}, which is still a high-dimensional optimization problem. We use the block coordinate descent algorithm (Tseng 2001) to solve \eqref{pro}. In each iteration we only optimize one run in the design. See Algorithm \ref{ag:cd} for detailed steps. This algorithm can also be used to construct approximate optimal designs with other criteria such as the minimax Wasserstein distance design. Similar strategy has been used to construction designs in the Euclidean space (Mu and Xiong 2017).

The initial points in Algorithm \ref{ag:cd} can be randomly generated. We begin with a relatively small $m$ and many initial points to get a maximin design $\mathcal{D}^{(m-1)}$ in $\widetilde{\mathcal{P}}_{q,p,m-1}([0,1],\tau)$. Consequently, $\mathcal{D}^{(m-1)}$ can be acted as the initial design in Algorithm \ref{ag:cd} to solve a maximin design $\mathcal{D}^{(2(m-1))}$ in $\widetilde{\mathcal{P}}_{q,p,2(m-1)}([0,1],\tau)$, which can better approximate the maximin design in $\mathcal{P}_{q,p}([0,1],\tau)$. For example, set the initial $m=11$, and repeat the above process twice. We successively obtain $\mathcal{D}^{(10)}$, $\mathcal{D}^{(20)}$, and $\mathcal{D}^{(40)}$. The design $\mathcal{D}^{(40)}$ is set as the final approximation to the maximin design in $\mathcal{P}_{q,p}([0,1],\tau)$. Figure \ref{fig:n6} shows the maximin $W_{q,p}$ distance designs for $n=6$ with $p=q=2$ and $p=q=1$ constructed via this way.

\begin{figure}[t]
	\begin{center}
		\scalebox{0.6}[0.6]{\includegraphics{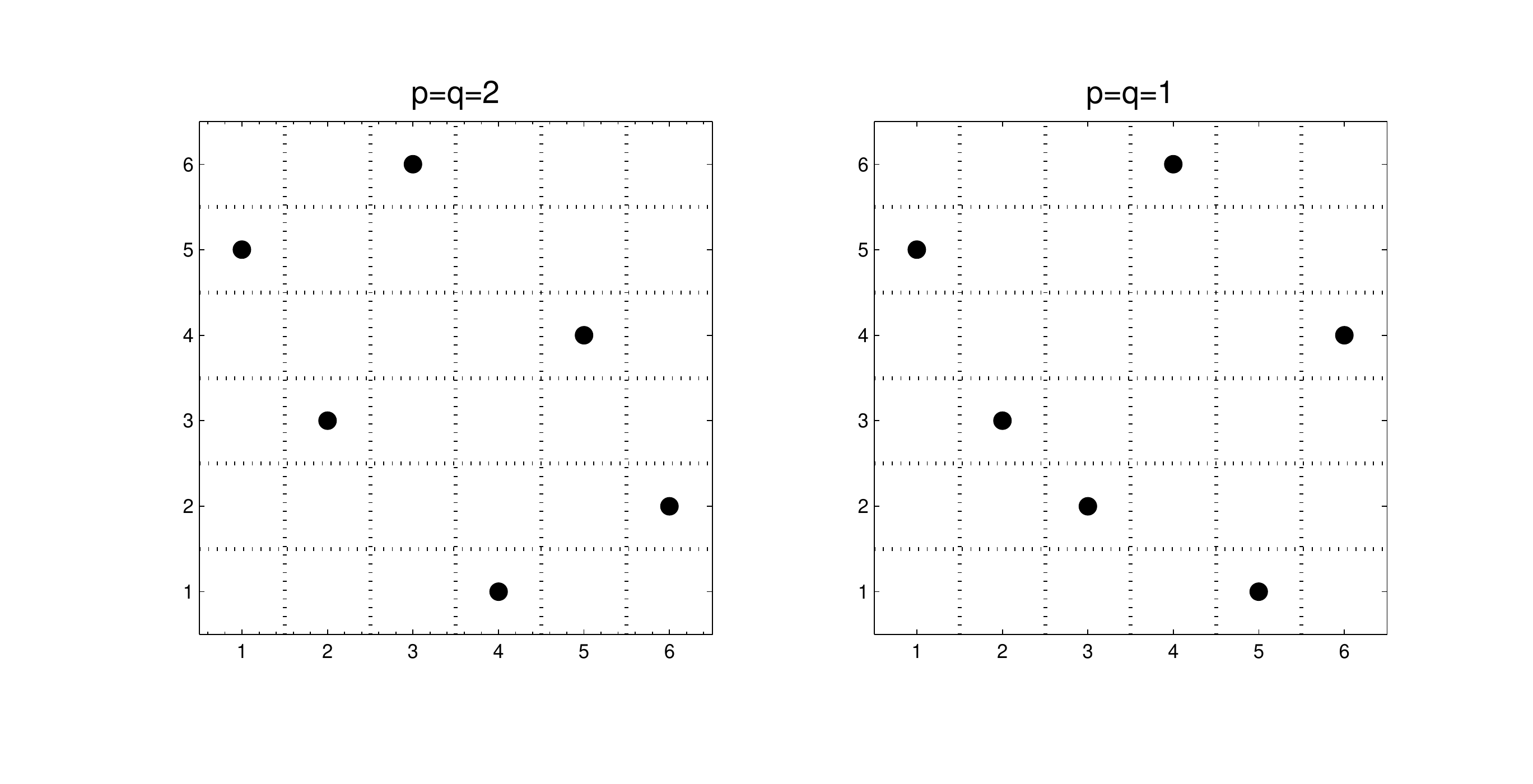}}
	\end{center}
	\caption{Maximin $W_{q,p}$ distance LH-type designs in $[0,1]\times\mathcal{P}_{q,p}([0,1],3)$ for $n=6$. The x-axis and y-axis denote indices of the base designs in the spaces $[0,1]$ and $\mathcal{P}_{q,p}([0,1],3)$, respectively.}\label{fig:lhn6}
\end{figure}

\subsection{LH-type designs in $[0,1]^d\times\mathcal{P}_{q,p}([0,1],\tau)$}\label{subsec:sdw}

When returning to the product space $[0,1]^d\times\mathcal{P}_{q,p}([0,1],\tau)$, we can similarly define the corresponding Wasserstein distance-based designs in this space. However, it is more difficult to find the optimal design with respect to the Wasserstein distance-based criteria. Here we propose a class of LH-type designs that possess good projection properties over both $[0,1]^d$ and $\mathcal{P}_{q,p}([0,1],\tau)$. Furthermore, we can find the optimal design over such LH-type designs, and the corresponding computation is relatively easier than over the whole space $[0,1]^d\times\mathcal{P}_{q,p}([0,1],\tau)$.

For an integer $n$, let $\mathcal{D}_1=\{\v{x}_1,\ldots,\v{x}_n\}$ and $\mathcal{D}_2=\{\mu_1,\ldots,\mu_n\}$ be space-filling designs in $[0,1]^d$ and $\mathcal{P}_{q,p}([0,1],\tau)$, respectively. For a permutation $(e_1,\ldots,e_n)$ of $(1,\ldots,n)$, let $\mathcal{D}(e_1,\ldots,e_n)=\{(\v{x}_1,\mu_{e_1}),\ldots,(\v{x}_n,\mu_{e_n})\}\subset[0,1]^d\times\mathcal{P}_{q,p}([0,1],\tau)$. The projections of $\mathcal{D}(e_1,\ldots,e_n)$ on $[0,1]^d$ and $\mathcal{P}_{q,p}([0,1],\tau)$ are respectively $\mathcal{D}_1$ and $\mathcal{D}_2$, which are both space-filling. Therefore, $\mathcal{D}(e_1,\ldots,e_n)$ can be viewed as an LH-type design (Mckay, Beckman, and Conover 1979). The designs $\mathcal{D}_1$ and $\mathcal{D}_2$ can be called base designs of $\mathcal{D}(e_1,\ldots,e_n)$. We can set $\mathcal{D}_2$ as the maximin Wasserstein distance design proposed in the previous subsection. For $\mathcal{D}_1$, there are many feasible choices (Joseph 2016; Santner, Williams, and Notz 2018). Here we prefer to those with good projection properties such as the maximin LH design (Park 1994), the maximum projection design (Joseph, Gul, and Ba 2015; Mu and Xiong 2018), and the rotated sphere packing design (He 2017), which usually yield good prediction of the simulation output.

Similar to optimal LH designs, we define the maximin Wasserstein distance LH-type design based on $(\mathcal{D}_1,\mathcal{D}_2)$ is the set $\mathcal{D}^*=\mathcal{D}(e^*_1,\ldots,e^*_n)=\{(\v{x}_1,\mu_{e^*_1}),\ldots,(\v{x}_n,\mu_{e^*_n})\}$, where $e^*_1,\ldots,e^*_n$ maximizes
\begin{equation}\label{mmc}   \mathrm{mdc}(\mathcal{D}(e_1,\ldots,e_n))=\min_{1\leqslant i<j\leqslant n}W_{q,p}(\delta_{\v{x}_i}\times\mu_{e_i},\delta_{\v{x}_j}\times\mu_{e_j})\end{equation}over all permutations $(e_1,\ldots,e_n)$ of $(1,\ldots,n)$. The Wasserstein distance in \eqref{mmc} can be computed by Theorem \ref{th:wf}, especially by \eqref{w11} and \eqref{w22}. Usually it is infeasible to compute the global solution to maximize (\ref{mmc})
based on all the $n!$ permutations. We can use the Monte Carlo method to approximate it by generating a large number of random permutations.
Figure \ref{fig:lhn6} shows the maximin $W_{2,2}$ distance LH-type design and the maximin $W_{1,1}$ distance LH-type design in $[0,1]\times\mathcal{P}_{q,p}([0,1],3)$ for $n=6$, where the base designs are the uniformly scattered points $\{0,\,1/5,\,2/5,\ldots,1\}$ and the corresponding maximin Wasserstein distance designs in Figure \ref{fig:n6}. Here the measures plotted with the red solid line, red dotted line, blue solid line, blue dotted line, green solid line, and green dotted line in Figure \ref{fig:n6} are denoted by $\mu_1,\ldots,\mu_6$. From Figure \ref{fig:lhn6} we can see that the optimal permutations in the two LH-type designs are $(5,3,6,1,4,2)$ and $(5,3,2,6,1,4)$, respectively.

\section{Gaussian process modeling}\label{sec:gp}

This section builds a Gaussian process model for computer experiments with both numeral and distribution inputs. We first define $h\circ\mu$ to be the probability measure of the random variable $h(X)$ for a function $h:\ \mathbb{R}\mapsto\mathbb{R}$, where $X\sim\mu$. For $\v{x}\in[0,1]^d$ and $\mu\in\mathcal{P}_{2,2}(\mathbb{R})$, we model the output of such a computer simulation as \begin{equation}f(\v{x},\mu)=\v{g}(\v{x})'\vg{\beta}+\int\alpha(t)d(h\circ\mu (t))+Z(\v{x},\mu),\label{kriging}\end{equation} where $\v{g}(\cdot)=\left(g_1(\cdot),\ldots,g_s(\cdot)\right)'$ and $h(\cdot)$ are
pre-specified functions, $\vg{\beta}$ is a vector of unknown regression coefficients, $\alpha(\cdot)$ is an unknown smooth function, and $Z(\v{x},\mu)$ is a stationary
Gaussian process defined on $[0,1]^d\times\mathcal{P}_{2,2}(\mathbb{R})$ with mean zero, variance $\sigma^2$, and covariance structure given below. The covariance between
$Z(\v{x}_1,\mu_1)$ and $Z(\v{x}_2,\mu_2)$ in \eqref{kriging} is represented by
\begin{equation*}\C[Z(\v{x}_1,\mu_1),
Z(\v{x}_2,\mu_2)]=\sigma^2R(\v{x}_1,\v{x}_2;\,\mu_1,\mu_2\,|\,\vg{\theta}), \end{equation*} where $R$ is the Gaussian correlation function
\begin{eqnarray}
R(\v{x}_1,\v{x}_2;\,\mu_1,\mu_2\,|\,\vg{\theta})=\exp\left\{-\sum_{i=1}^d\theta_i (x_{1i}-x_{2i})^2-\theta_{d+1}W_{2,2}(\mu_1,\mu_2)^2\right\},\label{GR}
\end{eqnarray} with positive correlation parameters $\vg{\theta}=(\theta_1,\ldots,\theta_d,\theta_{d+1})'$. Note that $R_W(\mu_1,\mu_2)=\\\exp\left\{-\theta_{d+1}W_{2,2}(\mu_1,\mu_2)^2\right\}$ gives a valid correlation structure of Gaussian processes on $\mathcal{P}_{2,2}(\mathbb{R})$ (Bachoc, Gamboa, Loubes, et al. 2018; Bachoc, Suvorikova, Ginsbourger, et al. 2020). The correlation structure in \eqref{GR} is valid for Gaussian processes on the product space of $[0,1]^d$ and $\mathcal{P}_{2,2}(\mathbb{R})$.

Since the infinity-dimensional $\alpha(\cdot)$ in \eqref{kriging} is hard to estimate, we use a linear expression to parameterize it. Let \begin{equation}\alpha(\cdot)=\vg{\gamma}'\v{b}(\cdot)\label{ib}\end{equation} with $\vg{\gamma}=(\gamma_1,\ldots,\gamma_l)'\in{\mathbb{R}}^l$, where $\v{b}(\cdot)=(b_1(\cdot),\ldots,b_l(\cdot))'$ are pre-specified basis functions. Then \eqref{kriging} can be approximated as
\begin{equation}f(\v{x},\mu)=\v{g}(\v{x})'\vg{\beta}+\left\{\int\v{b}(t)d(h\circ\mu (t))\right\}'\vg{\gamma}+Z(\v{x},\mu).\label{kriginga}\end{equation}

The parameters in model \eqref{kriginga} can be estimated by the maximum likelihood method. Suppose the set of input values is $\{(\v{x}_1,\mu_1),\ldots,(\v{x}_n,\mu_u)\}\subset[0,1]^d\times\mathcal{P}_{2,2}(\mathbb{R})$. The corresponding response values are $\v{y}=(f(\v{x}_1,\mu_1),\ldots,f(\v{x}_n,\mu_n))'$. The negative log-likelihood, up to an additive constant, is proportional to
\begin{equation}n\log(\sigma^2)+\log({\mathrm{det}}(\m{R}))+(\v{y}-\m{G}\vg{\beta}-\m{J}\vg{\gamma})'\m{R}^{-1}(\v{y}-\m{G}\vg{\beta}-\m{J}\vg{\gamma})/\sigma^2,\label{ll}\end{equation}
where $\m{R}$ is the $n\times n$ correlation matrix whose $(i,j)$th entry is $R(\v{x}_i,\v{x}_j;\,\mu_i,\mu_j\,|\,\vg{\theta})$ defined in (\ref{GR}), ``${\mathrm{det}}$" denotes
matrix determinant, $\m{G}=\left(\v{g}(\v{x}_1),\ldots,\v{g}(\v{x}_n)\right)'$, \\and $\m{J}=\left(\int\v{b}(t)d(h\circ\mu_1(t)),\ldots,\int\v{b}(t)d(h\circ\mu_n(t))\right)'$.
Denote $\m{U}=(\m{G}\ \m{J})$ and $\vg{\psi}=(\vg{\beta}'\ \vg{\gamma}')'$.

When $\vg{\theta}$ is known, the maximum likelihood estimators (MLEs) of $\vg{\psi}$ and $\sigma^2$ are
\begin{equation}\left\{\begin{array}{l}\hat{\vg{\psi}}=(\m{U}'\m{R}^{-1}\m{U})^{-1}\m{U}'\m{R}^{-1}\v{y},
\\\hat{\sigma}^2=(\v{y}-\m{U}\hat{\vg{\psi}})'\m{R}^{-1}(\v{y}-\m{U}\hat{\vg{\psi}})/n.\end{array}\right. \label{bs}\end{equation}
For an untried point $(\v{x}_0,\mu_0)\in[0,1]\times\mathcal{P}_{q,p}(\mathbb{R})$, the best linear unbiased predictor $\hat{f}$ of $f$ (Santner, Williams, and Notz 2018) is
\begin{eqnarray}\hat{f}(\v{x}_0,\mu_0)=\v{g}(\v{x}_0)'\hat{\vg{\beta}}+\left\{\int\v{b}(t)d(h\circ\mu_0(t))\right\}'\hat{\vg{\gamma}}+{\v{r}_0}'{\m{R}}^{-1}\big(\v{y}-\m{U}\hat{\vg{\psi}}\big),\label{blup}\end{eqnarray}
where ${\v{r}_0}=\big(R(\v{x}_0,\v{x}_1;\,\mu_0,\mu_1\,|\,{\vg{\theta}}),\ldots,R(\v{x}_0,\v{x}_n;\,\mu_0,\mu_n\,|\,{\vg{\theta}})\big)'$. Clearly this predictor possesses the interpolation property.
Given $\kappa\in(0,1)$, the $100(1-\kappa)\%$ prediction interval of $f(\v{x}_0,\mu_0)$ is\begin{equation}\label{ci}P\left(f(\v{x}_0,\mu_0)\in \hat{f}(\v{x}_0,\mu_0)\pm \eta(\v{x}_0,\mu_0)\,t_{n-s-l}(\kappa/2)\right)=1-a,\end{equation} where $\tau(\v{x}_0,\mu_0)\geqslant0$,
\begin{eqnarray*}&&\eta(\v{x}_0,\mu_0)^2=\frac{Q^2}{n-s-l}\\&&\quad\cdot\left\{1-\left(\v{g}(\v{x}_0)',\ \int\v{b}(t)'d(h\circ\mu_0 (t)),\ \v{r}_0'\right)\left(\begin{array}{cc}\m{0}&\m{U}'\\\m{U}&\m{R}\end{array}\right)^{-1}
\left(\begin{array}{c}\v{g}(\v{x}_0)\\\int\v{b}(t)d(h\circ\mu_0 (t))\\\v{r}_0\end{array}\right)\right\},\\&&Q^2=\v{y}'\big[\m{R}^{-1}-\m{R}^{-1}\m{U}(\m{U}'\m{R}^{-1}\m{U})^{-1}\m{U}'\m{R}^{-1}\big]\v{y},\end{eqnarray*}
and $t_{n-s-l}(\kappa/2)$ is the upper $\kappa/2$ quantile of the Student's $t$-distribution with $n-s-l$ degrees of freedom.
When $\vg{\theta}$ is unknown, by plugging \eqref{bs} into \eqref{ll}, we have the MLE of $\vg{\theta}$
\begin{equation*}\hat{\vg{\theta}}=\arg\min_{\vg{\theta}}\,n\log(\hat{\sigma}^2)+\log\left({\mathrm{det}}({\m{R}})\right).
\end{equation*}The predictor $\hat{f}$ in \eqref{blup} and the prediction interval in \eqref{ci} can be modified by replacing $\vg{\theta}$ with $\hat{\vg{\theta}}$.

Similar to the Gaussian process model on the Euclidean space in the literature, we call the above methods without and with the linear regression terms $\v{g}(\cdot)$ and $h(\cdot)$ in \eqref{kriging} simple Kriging and universal Kriging, respectively. Usually we set $h$ in \eqref{kriging} as the identity function,

There are some methods to parameterize $\alpha(\cdot)$ in \eqref{ib} such as the spline approximation (De Boor 1978). Here we propose the reconstruction parameterization approach (Xiong 2021) because of its good interpretation of the parameters. Specifically, in this approach the parameters $\gamma_i=\alpha(a_i),\ i=1,\ldots,l$, where $\{a_1,\ldots,a_l\}$ is the set of knots, and $\v{b}(\cdot)=(b_1(\cdot),\ldots,b_l(\cdot))'$ in \eqref{ib} are specific interpolation basis functions. When the distribution input in model \eqref{kriging} has a support $[0,1]$, we select $\v{b}$ as the polynomial interpolation basis functions (De Boor 1978), which have the Lagrange forms \begin{equation*}b_j(t)=\prod_{1\leqslant k\leqslant l,\ k\neq j}\frac{t-a_k}{a_j-a_k},\ \ j=1,\ldots,l.\end{equation*}We use the Chebyshev nodes \begin{equation*}\s{A}=\left\{a_j=1/2-\cos[(2j-1)\pi/2l]/2:\,j=1,\ldots,l\right\}\end{equation*} to avoid Runge's Phenomenon (De Boor 1978). The number of knots can be selected as $l=10$ according to the common $10d$ rule (Loeppky, Sacks, and Welch 2009).

When the distribution input $\mu$ in \eqref{kriging} lies in $\mathcal{P}_{q,p}([0,1],\tau)$, we use the design $\{\mu_1,\ldots,\mu_n\}\subset\widetilde{\mathcal{P}}_{q,p,m-1}([0,1],\tau)$ constructed in Section \ref{sec:sd}.
Let $h$ in \eqref{kriging} be the identity function. For $i=1,\ldots,n,\ j=1,\ldots,l$, by \eqref{tf} and \eqref{ap}, the entries of $\m{J}$ in \eqref{ll} can be given by
\begin{eqnarray*}&&\int b_j(t)\,d F_{\mu_i}(t)=\sum_{i=1}^{m-1}\int_{(i-1)/(m-1)}^{i/(m-1)}b_j(t)\,d F_{\mu_i}(t)\\&&=(m-1)\sum_{i=1}^{m-1}\left[F_{\mu_i}\left(\frac{i}{m-1}\right)-F_{\mu_i}\left(\frac{i-1}{m-1}\right)\right]\int_{(i-1)/(m-1)}^{i/(m-1)}b_j(t)\,d t.\end{eqnarray*}

\section{Numerical experiments with test functions}\label{sec:experiment}

\begin{figure}[t]
	\begin{center}
		\scalebox{0.6}[0.6]{\includegraphics{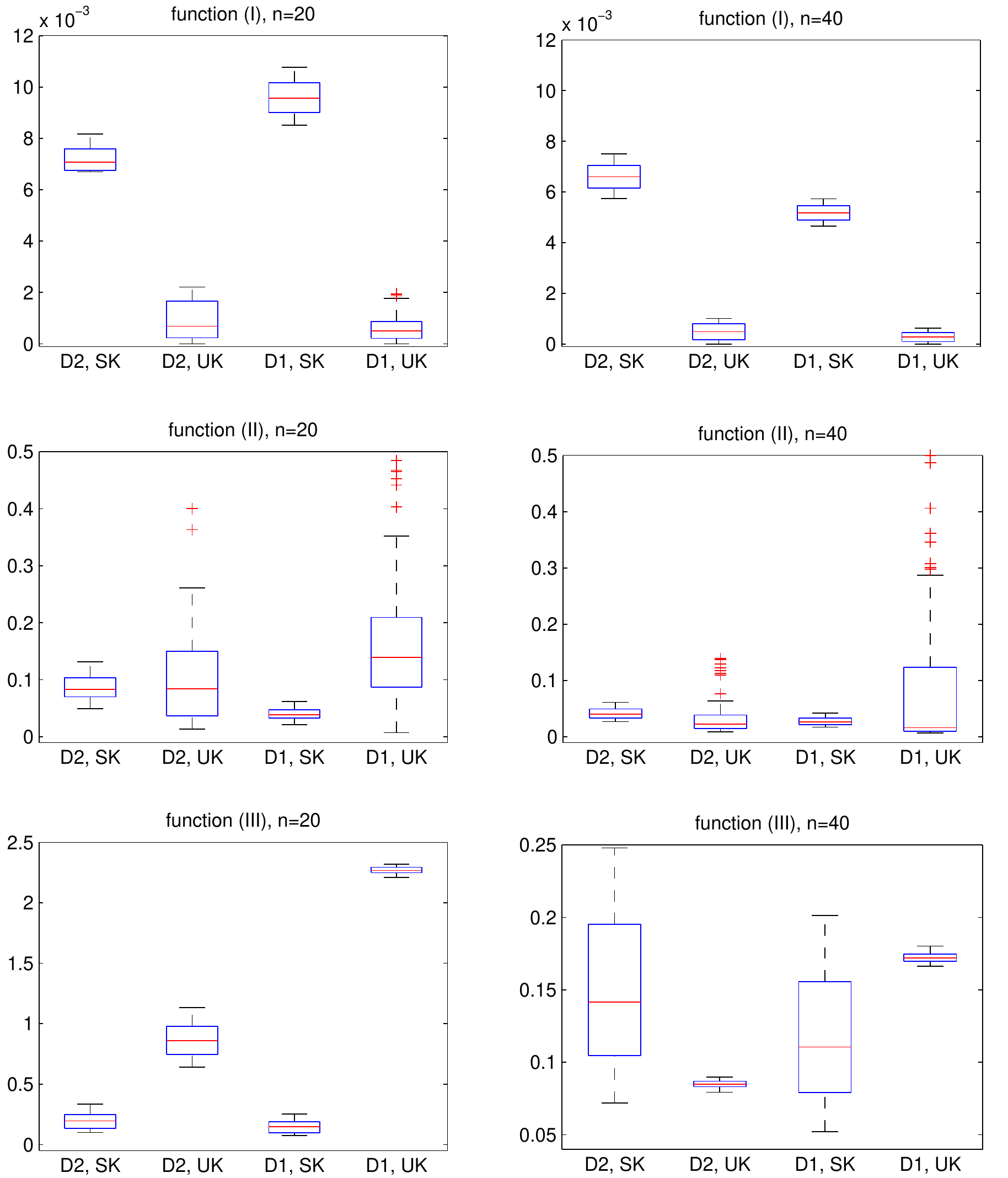}}
	\end{center}
	\caption{Box-plots of squared prediction errors.}\label{fig:box}
\end{figure}

In this section we conduct numerical experiments with the following test functions on $[0,1]^d\times\mathcal{P}_{q,p}([0,1],\tau)$,		
\begin{eqnarray*}&&\mathrm{(I)}\ f(x,\mu)=c+x^{1+c}+\int t\,d\mu(t),
\\&&\mathrm{(II)}\ f(x,\mu)=\int \cos(3t+c_1)\,d\mu(t)+\exp(x)+c_2F_\mu(x),		
\\&&\mathrm{(III)}\ f(x_1,x_2,\mu)=\left[x_1+\int t\,d\mu(t)+c_1\right]^2-c_2\log(1+x_2).\end{eqnarray*}The constants $c,\ c_1$, and $c_2$ in them are generated from the uniform distribution on $[0,1]$.
Four combinations of two design and two modeling methods are compared; see Figure \ref{fig:box}. D2 represents the maximin $W_{2,2}$-distance LH-type design based on $(\mathcal{D}_1,\mathcal{D}_2^{(2,2)})$ defined by \eqref{mmc}, where $\mathcal{D}_1$ is the maximin $\ell_2$-distance LH design and $\mathcal{D}_2^{(2,2)}$ is the maximin $W_{2,2}$-distance design. D1 represents the maximin $W_{1,1}$-distance LH-type design based on $(\mathcal{D}_1,\mathcal{D}_2^{(1,1)})$, where $\mathcal{D}_2^{(1,1)}$ is the maximin $W_{1,1}$-distance type design. Sample sizes in D2 and D1 are set as 20 and 40. Two modeling methods include simple Kriging (SK) and universal Kriging (UK). We use $\v{g}(\v{x})=\v{g}(x_1,\ldots,x_d)=(1,x_1,\ldots,x_d)'$ and $h(t)=t$ in the UK model \eqref{kriging}.

For each design and each test function, we generate training data, and then use SK and UK to build prediction models $\hat{f}$ in \eqref{blup}. The prediction accuracy is evaluated by the empirical squared prediction error, $\sum_{k=1}^{N}\big\{\hat{f}(\v{x}_k^*,\mu^*_k)-f(\v{x}_k^*,\mu^*_k)\big\}^2/N$ with $N=1000$, where the test data $(\v{x}_1^*,\mu^*_1),\ldots,(\v{x}_{N}^*,\mu^*_N)\in[0,1]^d\times\mathcal{P}_{q,p}([0,1],\tau)$ are generated randomly. Box-plots of the prediction errors with 100 repetitions are shown in Figure \ref{fig:box}. We can see that, with a relatively large $n$, UK is often better than SK, which is consistent with our empirical experience on Kriging in the Euclidean space. In practice, we can choose SK or UK through leave-one-out cross validation. It seems that, with SK, D1 is usually better than D2. Generally, D1 or D2 does not have clear superiority toward the other. Note that the computation for constructing D1 is more difficult than D2. We prefer to D2 in practice use.

\section{Applications to a metro simulation}\label{sec:real}

Urban metro systems are important components of urban transportation systems. The implementation of a metro system simulation provides a powerful instrument for system performance monitoring,
which enables operators to characterize the level of service and make decisions accordingly (Mo, Ma, Koutsopoulos, et al. 2021). Such a simulation is a typical stochastic simulation since a metro system contains a great deal of passengers' uncertainty, Here we consider a passenger flow simulation for a single metro route, and apply the proposed design and analysis methods to it.

On this route the passenger begins to tap in at the origin station and ends in tapping out at the destination station with taking only a train. The time between the arrival and departure of a passenger can be divided into access time, wait time, time on board, and egress time. Access time is the time it takes the passenger to walk from the tap-in fare gate to the platform; wait time is the time for which the passenger waits on the platform until boarding a train; and egress time is the time it takes to walk to the tap-out fare gate after alighting from the train.

\begin{figure}[t]
	\begin{center}
		\scalebox{0.6}[0.6]{\includegraphics{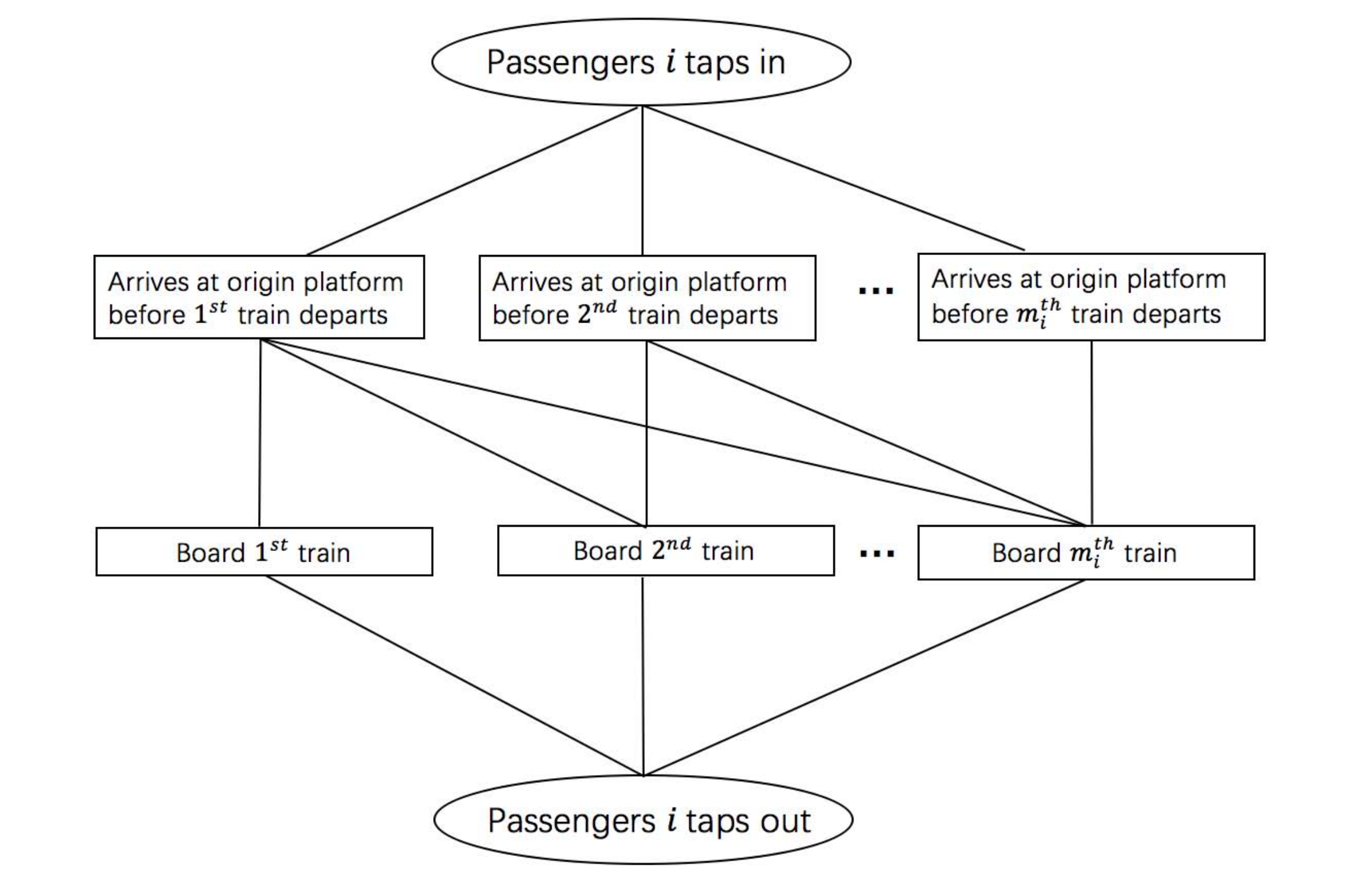}}
	\end{center}
	\caption{Flow chart of a passenger on a typical metro route}\label{fig:not}
\end{figure}
During peak hours, passengers may miss one or more trains due to crowded platform and carriages, and there may be more than one possible itineraries between their tap-in and tap-out times. Figure \ref{fig:not} illustrates all possible itineraries for passenger $i$ who enters and exits the metro system; a similar figure can be found in Zhu et al. (2017). We can use the passenger flow simulation to study the influence of the number of passengers, train schedule, and other factors on crowdedness, passenger-to-train assignment, and other responses we are interested in.

\begin{figure}[t]
	\begin{center}
		\scalebox{0.6}[0.6]{\includegraphics{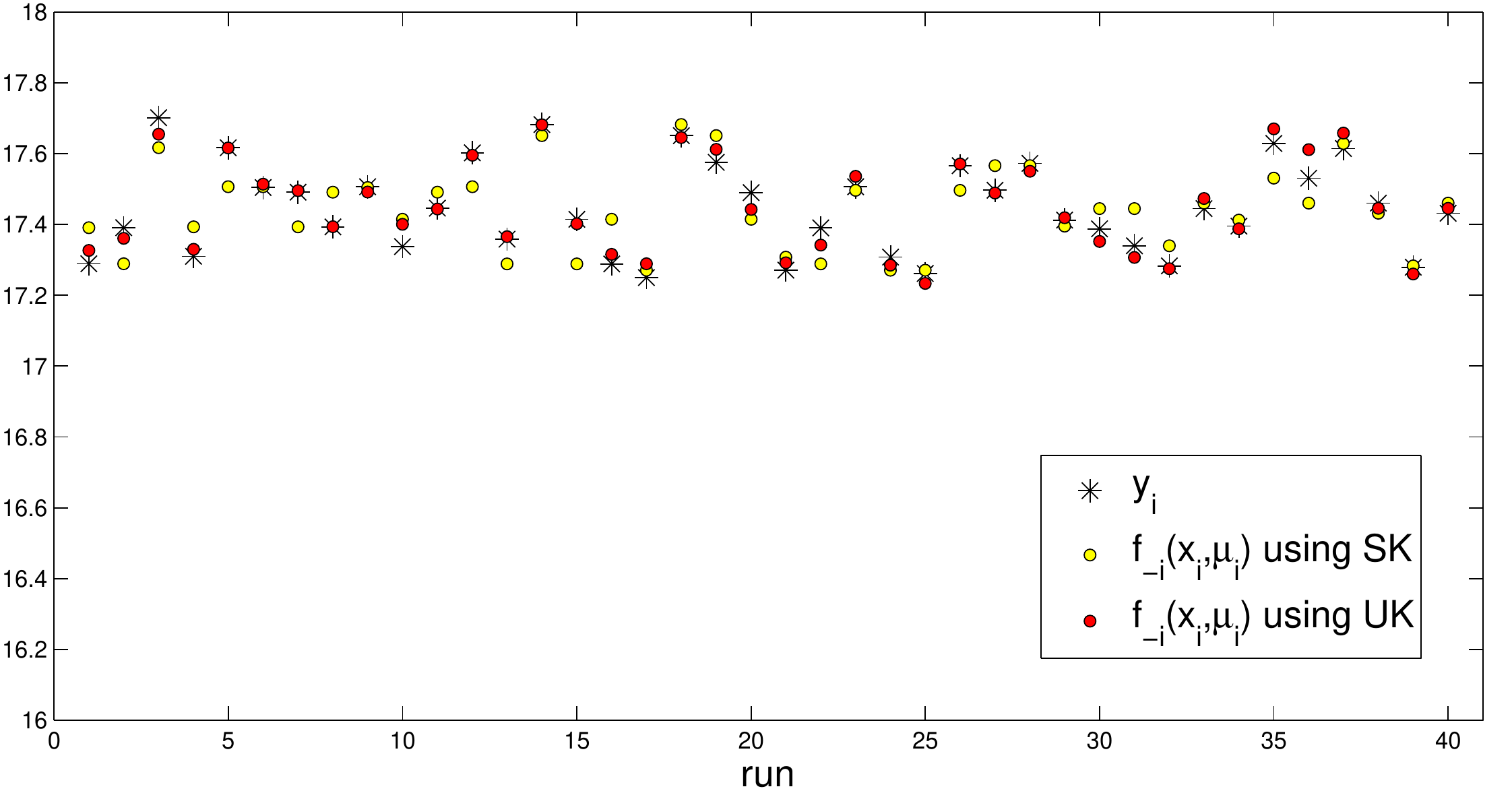}}
	\end{center}
	\caption{Leave-one-out prediction in Section \ref{sec:real}.}\label{fig:cv}
\end{figure}

We use the passenger flow simulator to conduct a one-day simulation. The inputs are as flows.
\begin{description}
\item (i) \emph{number of passengers in the day};
\item (ii) \emph{each passenger's origin and destination stations and tap-in time};
\item (iii) \emph{boarding probability of a passenger at each station (except the last station)};
\item (iv) \emph{probability distribution of a passenger's access time at each station (except the last station)};
\item (v) \emph{probability distribution of a passenger's egress time at each station (except the first station)};
\item (vi) \emph{train capacity};
\item (vii) \emph{train schedule including arrival and departure times at each station}.
\end{description}
The simulation can yield movement of each passenger in the metro system, including which train he/she takes, his/her tap-out time, and others.

Our simulation aims at simulating a real metro route of six stations. The inputs (i), (ii), (v), (vi), and (vii) can be obtained or estimated from the automatic fare collection system and automatic vehicle location system (Xiong et al. 2022; Li et al. 2022). The simulator designs the boarding probability for passenger $i$ in (iii) as \begin{equation}\label{bc}Pr(\rho;x)=\left\{\begin{array}{lll}0,&\quad\rho<1-x/2;\\(\rho-1+x/2)/x,&\quad\rho\in[1-x/2,1+x/2];\\1,&\quad\rho>1+x/2,\end{array}\right.\end{equation}where $\rho=(L-N_0)/N$, $L$ denotes the train capacity, $N_0$ denotes the current number of passerngers on this train, $N$ denotes the number of passengers who get on the platform earlier than passenger $i$ at the same platform, and $x\in[0,1]$ is a tuning parameter. We focus on the response surface $y=f(x,\mu)$, where the response $y$ is the mean travel time of passengers who tap in at the first station, $x\in[0,1]$ is the tuning parameter in \eqref{bc} at the first station, and $\mu\in\mathcal{P}_{2,2}([1,2],3)$ represents the distribution of access time at the first station. Other inputs of the simulator are fixed. The simulation contains $72,000$ passengers in the day.

\begin{figure}[t]
	\begin{center}
		\scalebox{0.6}[0.6]{\includegraphics{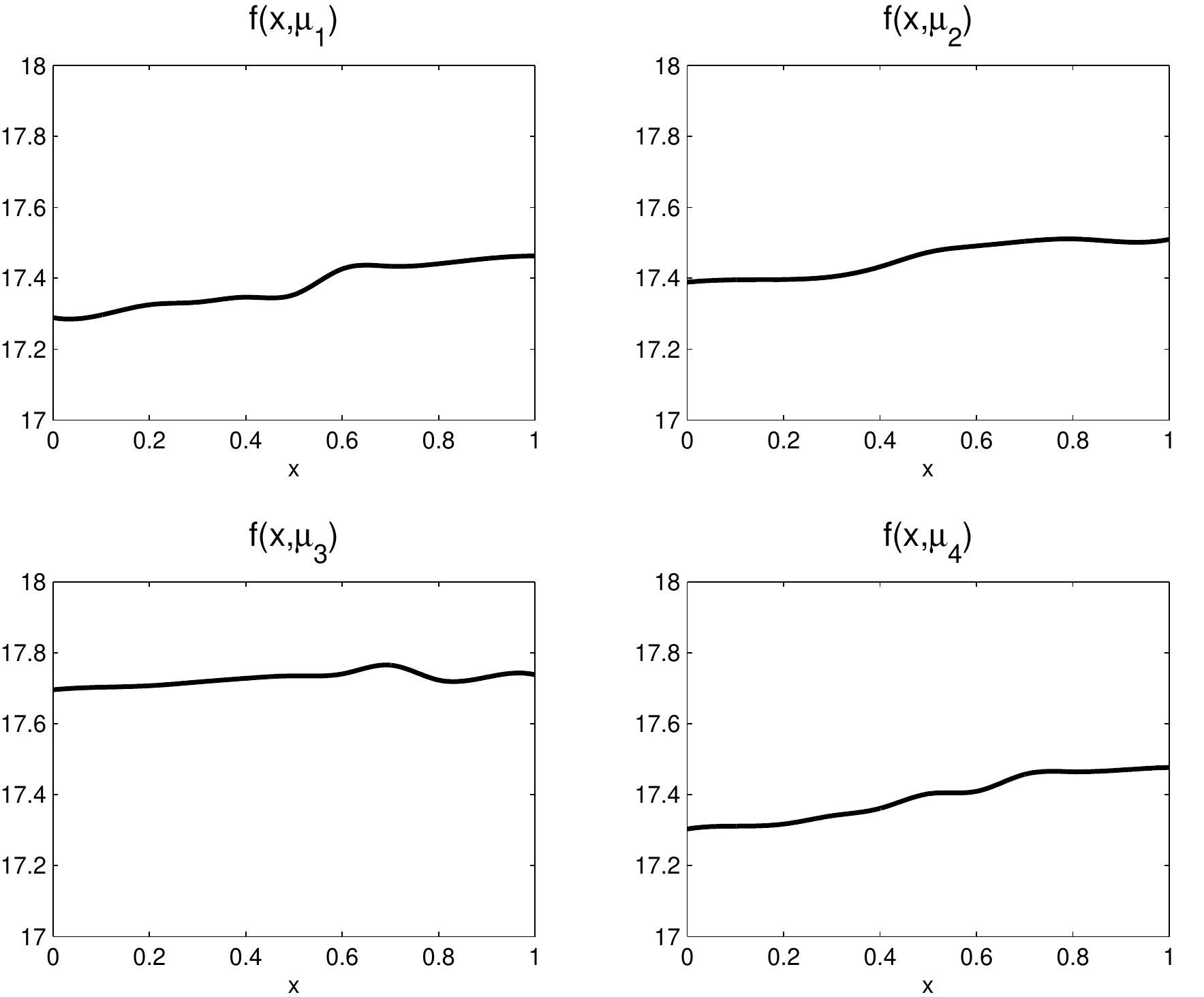}}
	\end{center}
	\caption{Profile curves of $f(x,\mu)$ with several fixed $\mu$ in Section \ref{sec:real}.}\label{fig:fixmu}
\end{figure}

We first use a 40-run maximin $W_{2,2}$-distance LH-type design which is the same as D2 in Section \ref{sec:experiment} to compute the corresponding values of $y$. Due to the randomness of the simulation, each value is obtained with 10,000 replicates. We then use SK and UK to build prediction models for the response. The true values and leave-one-out prediction values of $y$ are presented in Figure \ref{fig:cv}. It can be seen that UK is better than SK in most runs. On the numeral aspect, SK and UK yield leave-one-out mean squared prediction errors $4.8\times10^{-3}$ and $8.8\times10^{-4}$, respectively.

Consequently, we adopt the response surface constructed by UK based on all the data. Four profile curves of $f(x,\mu)$ with fixed $\mu$ are shown in Figure \ref{fig:fixmu}. We can see that, as the tuning parameter $x$ in their boarding probability increases, the travel times of the passengers from the first station have an growing trend. In the future we can develop sensitivity analysis methods to quantify the influence of the two input parameters $x$ and $\mu$ on the response. Another important issue is to calibrate the unobserved parameter $x$ with real data from the automatic fare collection system.

\section{Discussion}\label{sec:dis}

In this paper we have proposed design and modeling methods for computer experiments with both numeral and distribution inputs. We use the Wasserstein distance to unify the mixed inputs, and then the proposed methods can be viewed as straightforward extensions of the conventional methods for the Euclidean space. This makes our methods easy to understand and to implement.

There are several further topics we can follow in the future. It may be interesting to extend discrepancy-based (Fang, Lin, Winker et al. 2000) and lattice-based (He 2021) methods to construct space-filling designs in the probability measure space. Designs defined by other distribution distances or model-based designs can also be considered. Methods for sensitivity analysis, parameter calibration, and response optimization with both numeral and distribution inputs can be developed. In addition, possible directions include the study of distribution-output and/or multidimensional distribution-input computer simulations based on the Wasserstein distance, which calls for strategies to overcome the difficulties in computation.

\section*{Appendix: Proofs}

\emph{Proof of Lemma \ref{lemma:ei}}: It is obviously followed from the fact that any coupling measure in this case is  supported on the point $(\v{x},\v{y})$. In fact, this lemma has been appeared in literature, e.g.
line 2 on page 99 in Villani (2009). \qed

\noindent
\emph{Proof of Theorem \ref{th:wf}}: By the definition, \begin{eqnarray*}\mathcal{T}_{q,p}(\delta_{\v{x}}\times\mu,\delta_{\v{y}}\times\nu)=\inf_{\pi\in\Pi(\delta_{\v{x}}\times\mu,\delta_{\v{y}}\times\nu)}\int(\|\v{u}-\v{v}\|_p^p+|s-t|^p)^{\frac{q}{p}}d\pi(\v{u},s,\v{v},t).
\end{eqnarray*}

Since for any $\pi \in\Pi(\delta_{\v{x}}\times\mu,\delta_{\v{y}}\times\nu)$, the support lies in the fiber $\{(\v{x}, \cdot, \v{y}, \cdot)\}$,
\begin{eqnarray*}
\mathcal{T}_{q,p}(\delta_{\v{x}}\times\mu,\delta_{\v{y}}\times\nu)=\inf_{\pi^{\prime}\in\Pi (\mu,\nu)}\int(\|\v{x}-\v{y}\|_p^p+|s-t|^p)^{\frac{q}{p}}d\pi^{\prime}(s,t).
\end{eqnarray*}
Here, $\pi^{\prime}$ is the projection of $\pi$ along the first and third fibers. By this projection, two sets $\Pi(\delta_{\v{x}}\times\mu,\delta_{\v{y}}\times\nu)$ and  $\Pi(\mu, \nu)$ are in one-one correspondence.
We regard this problem as an optimal transportation from $\mathbb{R}$ to $\mathbb{R}$, with the cost function $c(t-s)$.  Recall that  $c(z) = (\|\v{x}-\v{y}\|_p^p+|z|^p)^{\frac{q}{p}}$, and simple calculation shows that it is convex. More precisely, $c$ is a convex nonnegative symmetric function. By Remark 2.19 (ii) in Villani (2015), we have

\begin{eqnarray*}
	\mathcal{T}_{q,p}(\delta_{\v{x}}\times\mu,\delta_{\v{y}}\times\nu)=\int_0^1  c(F_\mu^{-1}(t)-F_\nu^{-1}(t)) dt,
\end{eqnarray*}which completes the proof.
\qed

\section*{Acknowledgements}
Xiong's work is partially supported by National Key R\&D Program of China (Grant nos. 2021YFA1000300, 2021YFA1000301, and 2021YFA1000303) and the National Natural Science Foundation of China (Grant no. 12171462). Cui's work  is supported by the National Natural Science Foundation of China (Grant Nos. 12171234, 11790272), the Project Funded by the Priority Academic Program
Development of Jiangsu Higher Education Institutions (PAPD) and the Fundamental Research Funds for the Central Universities.

\vspace{1cm} \noindent{\Large\bf References}

{\begin{description}
\footnotesize
\item
Arjovsky, M., Chintala, S., and Bottou, L. (2017), Wasserstein generative adversarial networks, \textit{Proceedings of the 34th International Conference on Machine Learning}, 70, 214--223.

\item
Bachoc, F., Gamboa, F., Loubes, J.-M., and Venet, N. (2018), A Gaussian process regression model for distribution inputs, \textit{IEEE Transactions on Information Theory}, 64, 6620--6637.

\item
Bachoc, F., Suvorikova, A., Ginsbourger, D., Loubes, J.-M., and Spokoiny, V. (2020), Gaussian processes with multidimensional distribution inputs via optimal transport and Hilbertian
embedding, \textit{Electronic Journal of Statistics}, 14, 2742--2772.

\item
Betancourt, J., Bachoc, F., Kleina, T., Idier, D., Pedreros, R., Rohmer, J. (2020), Gaussian process metamodeling of functional-input code for coastal flood
hazard assessment, \textit{Reliability Engineering and System Safety}, 106870.

\item
Chen, J., Mak, S., Joseph, V. R., and Zhang, C. (2021), Function-on-function Kriging, with applications to three-dimensional printing of aortic tissues,
\textit{Technometrics}, 63, 384-395,

\item{}
De Boor, C. (1978), \textit{A Practical Guide to Splines}, Springer.

\item{}
Elefteriadou, L. (2014), \textit{An Introduction to Traffic Flow Theory}, Springer.

\item
Fang, K. T., Li, R. Z., and Sudjianto, A. (2005), \textit{Design and Modeling for Computer Experiments}, Chapman Hall/CRC Press.

\item
Fang, K. T, Lin, D. K. J., Winker, P., Zhang, Y. (2000), Uniform design: Theory and application. \textit{Technometrics}, 42, 237--248

\item
He, X. (2017), Rotated sphere packing designs, \textit{Journal of the American Statistical Association}, 112, 1612--1622.

\item
He, X. (2021), Lattice-based designs possessing quasi-optimal separation distance on all projections, \textit{Biometrika}, 108, 443--454.

\item
Johnson, M., Moore, L., and Ylvisaker, D. (1990), Minimax and maximin distance design, \textit{Journal of Statistical Planning and Inference}, 26, 131--148.

\item
Joseph, V. R. (2016), Space-filling designs for computer experiments: A review, \textit{Quality Engineering}, 28, 28--35.

\item
Joseph, V. R., Gul, E., and Ba, S. (2015), Maximum projection designs for computer experiments, \textit{Biometrika}, 102, 371--380.

\item{}
Loeppky, J. L, Sacks, J., and Welch, W. J. (2009). Choosing the sample size of a computer experiment: A practical guide, \textit{Technometrics}, 51, 366--376.

\item{}
Li, C., Xiong, S., Sun, X., Qin, Y. (2022), Bayesian analysis for metro passenger flows using automated data, \textit{Mathematical Problems in Engineering}, Article ID: 9925939.

\item{}
Mckay, M. D., Beckman, R. J., and Conover, W. J. (1979), A comparison of three methods for selecting values of input variables in the analysis of output from a computer code, \textit{Technometrics}, 21, 239--245.

\item{}
Morris, M.D. and Mitchell, T.J. (1995), Exploratory designs for computational experiments, \textit{Journal of Statistical Planning and Inference}, 43, 381--402.

\item
Mu, W. and Xiong, S. (2017), On algorithmic construction of maximin distance designs, \textit{Communications in Statistics: Simulation and Computation}, 46, 7972--7985.

\item
Mu, W. and Xiong, S. (2018), A class of space-filling designs and their projection properties, \textit{Statistics \& Probability Letters}, 141, 129--134.

\item
Mo, B., Ma, Z., Koutsopoulos, H. N., and Zhao, J. (2021), Calibrating path choices and train capacities for urban rail transit simulation models using smart card and train
movement data, \textit{Journal of Advanced Transportation}, Article ID: 5597130.

\item
Muehlenstaedt, T., Fruth, J., and Roustant, O. (2017), Computer experiments with functional inputs and scalar outputs
by a norm-based approach, \textit{Statistics and Computing}, 27, 1083--1097.

\item
Nanty, S., Helbert, C., Marrel, A., P\'{e}rot, N., and Prieur, C. (2016), Sampling, metamodeling, and sensitivity analysis of numerical simulators with functional stochastic inputs, \textit{SIAM/ASA Journal on Uncertainty Quantification}, 4, 636--659.

\item
Pamaretps, V. M. and Zemel, Y. (2020), \textit{An Invitation to Statistics in Wasserstein Space}. Springer.

\item{}
Park, J. S. (1994), Optimal Latin-hypercube designs for computer experiments, \textit{Journal of Statistical Planning and Inference}, 39, 95--111.

\item
Peyr\'{e}, G. and Cuturi, M. (2019), Computational optimal transport, \textit{Foundations and Trends in Machine Learning}, 11, 355--607.

\item{}
Santner, T. J., Williams, B. J., and Notz, W. I. (2018). \textit{The Design and Analysis of Computer Experiments}, 2nd Edition, Springer. 

\item{}
Squazzoni, F., Jager, W., and Edmonds, B. (2014), Social simulation in the social sciences: A brief overview, \textit{Social Science Computer Review}, 32, 279--294.

\item{}
Tan, M. H. Y. (2019), Gaussian process modeling of finite element models with functional inputs, \textit{SIAM/ASA Journal on Uncertainty Quantification}, 7, 1133--1161.

\item
Tseng, P. (2001), Convergence of a block coordinate descent method for nondifferentiable minimization, \textit{Journal of optimization theory and applications}, 109, 475--494.

\item
Villani, C. (2009), \textit{Optimal Transport, Old and new}, Springer.

\item
Villani, C. (2015), \textit{Topics in Optimal Transportation}, AMS.

\item
Xiong, S. (2021), The reconstruction approach: From interpolation to regression, \textit{Technometrics}, 63, 225--235,

\item
Xiong, S. Li, C. Sun, X. Qin, Y., and Wu, C. F. J. (2022), Statistical estimation in passenger-to-train assignment models based on automated data, \textit{Applied Stochastic Models in Business and Industry}, 38, 287--307.

\item
Xiu, D. (2010), \textit{Numerical Methods for Stochastic Computations: A Spectral Method Approach}. Princeton University Press.

\item
Zhu, Y., Koutsopoulos, H. N., and Wilson, N. H. M. (2017), A probabilistic passenger-to-train assignment model based on automated data, \textit{Transportation Research Part B: Methodological}, 104, 522--542.
\end{description}}

\end{document}